\documentclass[11pt,twoside]{article}
\usepackage{asp2010}

\resetcounters

\begin{document}

\title{First Results of The Konkoly Blazhko Survey II}
\author{\'A.~S\'odor$^1$, J.~Jurcsik$^1$, L.~Moln\'ar$^1$, B.~Szeidl$^1$, Zs.~Hurta$^1$, G.~\'A.~Bakos$^2$, J.~Hartman$^2$, B.~B\'eky$^2$, R.~W.~Noyes$^2$, D.~Sasselov$^2$, T.~Mazeh$^4$, J.~Bartus$^5$, B.~Belucz$^3$, G.~Hajdu$^6$, Zs.~K\H ov\'ari$^1$, E.~Kun$^7$, I.~Nagy$^3$, K.~Posztob\'anyi$^6$, P.~Smitola$^3$, K.~Vida$^1$
\affil{$^1$Konkoly Observatory of the Hungarian Academy of Sciences, P.O.~Box~67, H-1525 Budapest, Hungary}
\affil{$^2$Harvard-Smithsonian Center for Astrophysics, Cambridge, MA}
\affil{$^3$Dept. of Astronomy, E\"otv\"os University, 1518 Budapest PO Box 49, Hungary}
\affil{$^4$School of Physics and Astronomy, Raymond \& Beverly Sackler Faculty of Exact Sciences, Tel Aviv University, Tel Aviv 69978, Israel}
\affil{$^5$Astrophysical Institute Potsdam, An der Sternwarte 16, 14482 Potsdam, Germany}
\affil{$^6$Visiting astronomer at the Konkoly Observatory of the Hungarian Academy of Sciences}
\affil{$^7$University of Szeged, Dept. of Exp. Physics and Astron. Obs., 6720 Szeged, D\'om t\'er 9, Hungary}
}

\begin{abstract}

The two parts of the Konkoly Blazhko Survey (KBS~I and II) are introduced. The most important preliminary findings of the second part are presented in comparison to the results of the first part. Two interesting cases of very strong modulation from the KBS~II are also shown.

\end{abstract}

\section{The Survey}
The Konkoly Blazhko Survey (KBS) I and II aim to collect accurate, extended, multicolour light curves of bright, northern, fundamental mode RR Lyrae stars of the Galactic field \citep{KBS0,J_09_Bl} in order to determine the incidence rate of the modulation, to study the modulation properties in detail and to study temporal changes in the modulation properties.

The first, already finished part of the survey, KBS~I, was initiated in 2004. Altogether, 30 bright, short-period ($P < 0.5$\,d) variables were observed with the 60-cm automatic telescope of the Konkoly Observatory on about 750 nights. A surprisingly high number of stars (14 out of 30; 47\%) were found to show the Blazhko effect. The most important findings of KBS~I has already been summarized in \citet{KBS1}.

In 2009, the second part of the survey, KBS~II, was launched to obtain a sample of longer-period RRab stars. Our main goal was to check whether the frequent occurrence of the modulation found in the KBS~I sample is a general property of RRab stars, or the ratio is pulsation-period dependent. In addition to the multicolour observations made with the telescopes of the Konkoly Observatory, Hungarian-made Automated Telescope Network (HATNet) observations and light-curve data from public data bases (ASAS -- \citealt{asas}, NSVS -- \citealt{nsvs}, WASP -- \citealt{wasp}) were also used to study the whole sample of known RRab stars matching the following criteria: declination above +10 deg, 2MASS $K$ magnitude brighter than 13.0 mag, pulsation period in the 0.55--0.60\,d range. The sample of KBS~II consists of 124 objects.

While the moderate sample size of KBS~I allowed a deep study of those variables, the much larger KBS~II sample was better suited for statistical analysis of the modulation properties.

\section{Light-Curve Classification}

The objects were classified into three groups based on the character of the light curves: (1) stable light curve, (2) modulated light curve or (3) insufficient information.

First, traces of modulation were sought. If at least one dataset of an object showed definite variations in maximum brightness, or the Fourier spectrum contained 
modulation peaks around the pulsation harmonics, the object was classified as modulated. If no modulation was found, we investigated the data quality to decide wether the light curve can be classified as stable or the available information was insufficient either to detect or exclude modulation.

About one-third (45) of the investigated objects showed modulation, while 60 objects had stable light curve. The remaining 19 objects of the sample had insufficient observational data for establishing light-curve stability or modulation.

\section{Results and Comparison with KBS~I}

The incidence rate of the modulation in the KBS~II sample, disregarding the 19 uncertain objects, is 43\%. The exact ratio depends on which objects have been judged as uncertain, though. The modulation rate found is only slightly below the value we have found among the short-period field RR Lyrae stars of the KBS~I sample (47\%). This difference can be explained with the lower pulsation and modulation amplitudes of the longer-period RR Lyrae stars \citep{J_modamp}. It should be emphasized that the found incidence rate is a lower limit only, since there could be further Blazhko stars with modulation amplitudes below the detection limit or the modulation can be temporal.

In the KBS~II sample, the phase relation between the amplitude and phase variations of the maximum light is such that, in most cases, the light maximum travels around a counterclockwise loop during the modulation. In other words, the earliest-maximum phase precedes the lowest-amplitude phase. This direction is clockwise only in one case, counterclockwise at 20 objects and the loop is degenerated or uncertain at the rest of the Blazhko stars. Similarly, in the KBS~I sample, only one object, UZ~Vir has clockwise direction. This loop direction is related to the asymmetry of the modulation peaks \citep{SZJ_mod}.

\section{Two Very Similar and Strong Modulations}

\begin{figure}[!ht]
  \begin{center}
     \includegraphics[height=80mm]{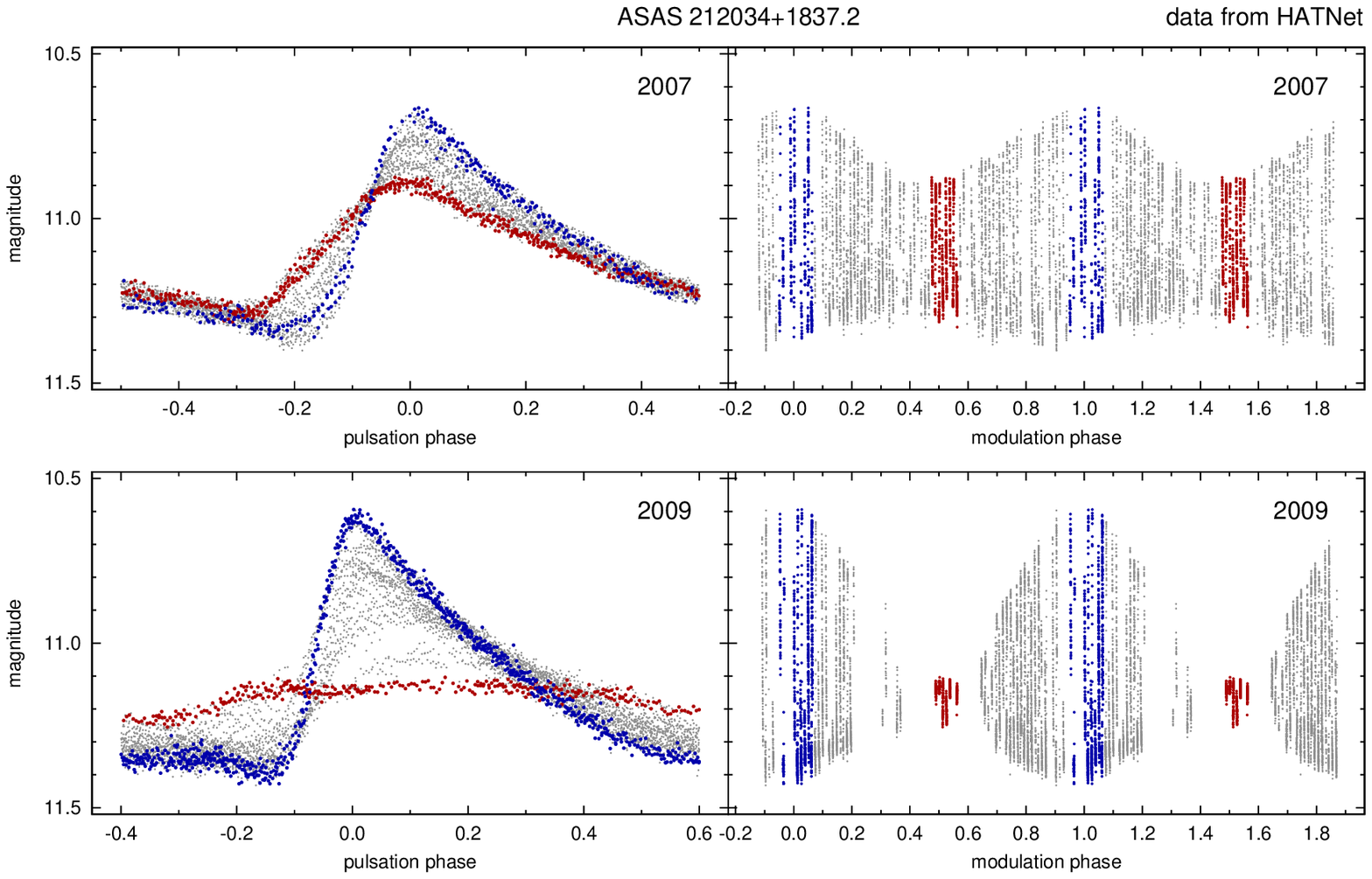}
     \includegraphics[height=80mm]{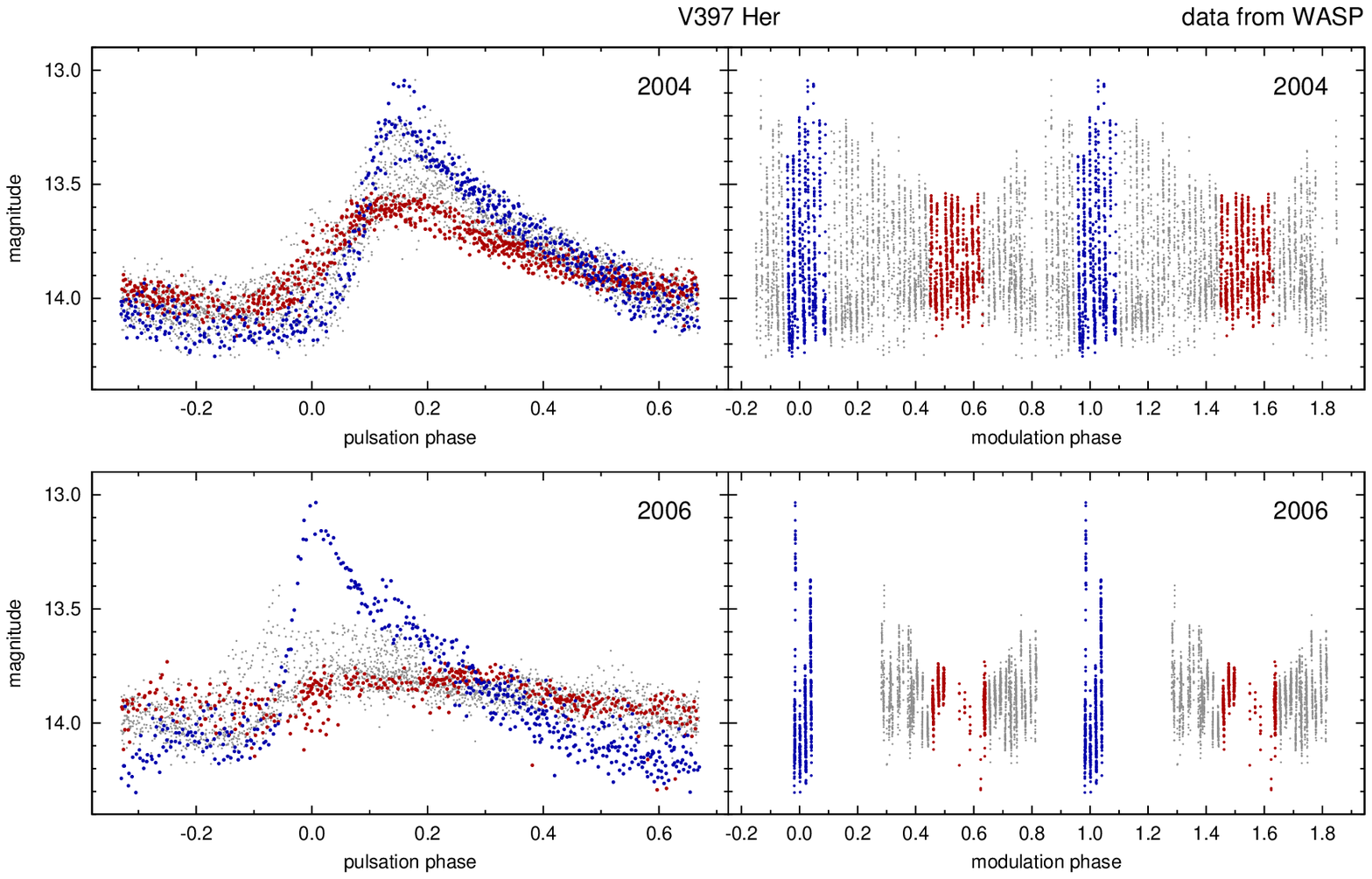}
   \end{center}
   \label{fig:theonlyonefigure}
 \caption{The Blazhko effect of two strongly modulated targets of KBS~II in different seasons. The highest- and lowest-amplitude Blazhko phases are marked with different shades/colours. The strength of the modulation changed in both cases similarly during the nearly two years elapsed between their two observing seasons.}
\end{figure}

Two of the Blazhko stars in the KBS~II sample exhibited so strong modulation that the pulsation almost disappeared at the lowest-amplitude phase. The light curves of these two objects, ASAS~212034+1837.2 (HATNet data) and V397~Her (WASP data) are shown in the panels of Fig.~\ref{fig:theonlyonefigure}. Similarly drastic light-curve variations have already been detected in V442~Her \citep{Sch_2000_V442Her}, V445~Lyr (KIC~6186029; \citealt{B_2010Kep}) and V18 of M5 \citep{J_M5b}.

A significant difference is, however, that in V442~Her and V18 of M5, no Blazhko period was found, only long-term pulsation amplitude variations were detected. Contrarily, in the two very strongly modulated objects of the KBS~II sample, definite Blazhko periods could be determined (82 d and 52 d for ASAS~212034+1837.2 and V397 Her, respectively).

It is stunning that these two objects, ASAS~212034+1837.2 and V397~Her, behaved so similarly, even though they both have complex modulation properties. Both of these objects were observed in two seasons and the very strong modulation occurred only in one of the seasons. Both stars' modulation was modestly strong in their respective first observing seasons (see upper panels of the objects). Two years later, both exhibited almost the strongest possible modulation (bottom panels). These changes in the modulation could be the result of multiperiodic modulation, where the two modulations caused beating and at certain phases they amplified each other similarly to the modulation of CZ~Lac in 2004 \citep{S_CZL}. Note that, according to Fig. 2 of \citet{B_2010Kep}, this might be the case in V445~Lyr, too.

Such a strong modulation can hardly be explained by non-radial modes superimposed on the fundamental radial pulsation mode \citep{J_M5b}.

\acknowledgements The financial support of the Hungarian OTKA grants T-068626 and K-081421 is acknowledged. HATNet operations have been funded by NASA grants NNG04GN74G, NNX08AF23G and SAO IR\&D grants.

\bibliography{sodor}

\end{document}